\documentclass[aip,jmp,amsmath,amssymb,reprint,]{revtex4-1}

\usepackage[pdftex]{graphicx}
\usepackage{dcolumn}
\usepackage{bm}

\begin{document}



\title{Direct measurement of plasmon propagation lengths on lithographically defined metallic waveguides on GaAs} 

\author{G. Bracher}
\email{gregor.bracher@wsi.tum.de}
\author{K. Schraml}
\author{C. Jakubeit}
\author{M. Kaniber}
\author{J. J. Finley}

\affiliation{Walter Schottky Institut and Physik Department, Technische Universit\"at M\"unchen, Am Coulombwall 4, 85748 Garching, Germany}

\date{\today}

\begin{abstract}

We present optical investigations of rectangular surface plasmon polariton waveguides lithographically defined on GaAs substrates. The plasmon propagation length is directly determined using a confocal microscope, with independent polarization control in both excitation and detection channels. Surface plasmon polaritons are launched along the waveguide using a lithographically defined defect at one end. At the remote end of the waveguide they scatter into the far-field, where they are imaged using a CCD camera. By monitoring the length dependence of the intensity of scattered light from the waveguide end, we directly extract the propagation length, obtaining values ranging from $L_{SPP}=10-40\mu m$ depending on the waveguide width ($w_{WG} =2-5~\mu m$) and excitation wavelength ($760-920~nm$).  Results are in good accord with theoretical expectations demonstrating the high quality of the lithographically defined structures.  The results obtained are of strong relevance for the development of future semiconductor based integrated plasmonic technologies. 

\end{abstract}


\maketitle 

%
%

The drive to miniaturize photonic devices to sizes far below the optical wavelength has fueled much interest in the surface plasmon polaritons (SPPs) \cite{{Barnes2003},{Maier2007}}, optically active charge density waves that are tightly bound at metal-dielectric interfaces. Such excitations support propagating modes even when their size is far below the free space optical wavelength ($\lambda_{free}$). A number of impressive demonstrations have already been made, including guiding of SPPs along sub-wavelength metallic waveguides with sizes\cite{Krenn2002} $<10\%$ of $\lambda_{free}/2$ and also sub-wavelength focusing of surface plasmon polaritons \cite{yin2005}.
The successful realization of integrated plasmonic technologies will undoubtedly necessitate \textit{hybrid}-approaches in which plasmonic components are integrated on-chip with sources, detectors\cite{falk2009near} and married with conventional electrical devices \cite{atwater2010plasmonics}.  This calls for the development of reliable lithographic fabrication techniques on \textit{semiconductor} substrates. First studies of such hybrid plasmonic devices have started to appear.  For example, surface attached semiconductor nano-crystals have been employed to deterministically generate SPPs in chemically synthesized metallic nanowires at the quantum limit\cite{akimov2007}. Moreover, integration of superconducting single photon detectors with plasmonic waveguides (WGs) have allowed the observation on non-classical plasmon statistics \cite{heeres2009}.
For future applications, especially in the field of quantum information processing, the use of near surface self-assembled  semiconductor quantum dots (QDs) coupled to surface plasmonic elements provides significant advantages, since the emitter is naturally integrated into the substrate at a position in the near field of the SPP mode.
Motivated by the potential to combine semiconductor quantum emitters with plasmonic nano-structures, we present in this article a systematic investigation of propagation along Au waveguide defined on a GaAs substrate. We demonstrate that plasmons can be launched into SPP waveguide modes with widths down to $2~\mu m$ by illuminating a lithographically defined defect\cite{ditlbacher2002a}, obviating the need for grating couplers for excitation\cite{Devaux2003}. After propagation over length scales $>15~\mu m$ SPPs are scattered into the far-field, where they are imaged using a CCD. By employing a cross-polarized excitation-detection geometry, we achieve a strong $(>10^4)$ suppression of the excitation laser allowing the direct measurement of SPP propagation lengths without employing spatial filters  in our optical setup. Polarization dependent measurements reveal that both the excitation and out-coupled light are strongly polarized, with degrees of linear polarization (DoP) $\geq 70 \%$. Maximum in-coupling is obtained for light polarized along the waveguide axis \cite{lamprecht2001}. Measurements of the propagation length are performed for different wire widths from $2-5~\mu m$ and for different wavelengths spanning the GaAs\cite{blakemore1982} bandgap ($760~nm$ - $920~nm$). The propagation length clearly increases with increasing wavelength and increasing waveguide width reflecting the modal structure of the guided SPPs. The investigated wavelength regime spans the typical emission energies of self-assembled InGaAs quantum dots \cite{kaniber2011}, demonstrating that the methods developed ideally lend themselves to future quantum optical experiments using active emitters\cite{{ambati2008},{Cuche2010}}

\begin{figure}[t!]
\includegraphics[width=0.95\columnwidth]{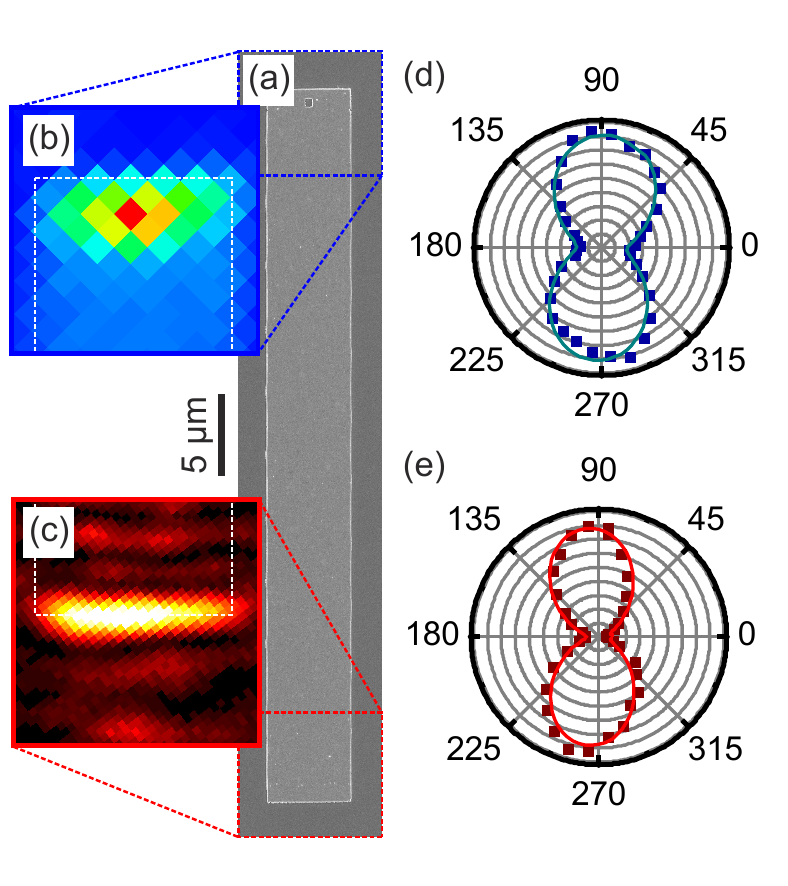}
\caption{\label{pol} (a) Scanning electron microscope image of a typical $5~\mu m$ wide and $35~\mu m$ long waveguide with a lithographically manufactured scattering hole. (b) a spatial scan of the excitation laser showing that  in-coupling is strongly enhanced at the scattering hole. (c) a map of the out-coupling: scattering to the far-field can be observed only at the end of the waveguide. (d) polarization dependence of the excitation: the electric field has to be oriented along the WG axis for an optimal in-coupling. (e) polarization dependence of the detection: the scattered photons are linearly polarized along the WG axis.}
\end{figure}

%
%

As discussed in the methods section, plasmonic waveguides were defined using electron beam lithography and deposition of a 100nm thick Au film on undoped GaAs substrates.
A scanning electron microscope (SEM) image of a typical WG is presented in figure \ref{pol} (a). The WG length varies from $L_{WG}=15$ to $45~\mu m$  in $5~\mu m$ steps and the WG width from $w_{WG}=2 $ to $5~\mu m$ in $1~\mu m$ steps. To enhance the in-coupling efficiency for light incident normal to the plane of the WG, a small defect is defined at one end of the WG. This defect is a rectangular 450x450nm hole. For optical characterization we used a confocal microscopy system that provides diffraction limited performance and facilitates independent control of polarization in both excitation and detection channels. Light from a tunable Ti:Sa laser is linearly polarized and adjusted using a $\lambda/2$-plate. It is then focused onto the sample via a 50:50 beam splitter using a 50x microscope objective with numerical aperture $NA=0.55$. Plasmons are generated that propagate along the WG axis before being scattered into the far field at the end and imaged using a Peltier cooled CCD camera. A second polarizer is placed in the detection channel using a cross-polarized geometry in order to suppress directly scattered light. Tilting the waveguide $45^\circ$ with respect to the excitation polarization provides the best trade off between in-coupling efficiency and detected signal while still providing a suppression of stray light from the excitation laser by a factor ($>10^4$). The SEM image in figure 1 is superimposed onto two spatial resolved measurements, that confirm that SPPs are generated in our experiments. Figure \ref{pol} (b) shows a spatially resolved image of the in-coupling obtained by raster scanning the excitation spot over the structure and measuring the intensity of the light emitted from the WG end. We observe a strong ($>5 \times$) enhancement of the in-coupling efficiency when the laser is focused directly onto the scattering hole. Furthermore, figure \ref{pol} (c) shows that the detected signal arises solely from the WG end. This observation indicates that the WGs are of high quality, since no scattering along the WG can be observed\cite{Flynn2010}. For a detailed analysis of the in- and out-coupling properties, the polarization dependence of the excitation and detection was investigated. Figure \ref{pol} (d) shows that the most efficient in-coupling is achieved when polarization of the excitation is aligned along the waveguide axis. The $w_{WG}= 5 \mu m$ broad WG sample presented in figure 1 exhibits a degree of polarization ($DoP = \frac{I_{max}-I_{min}}{I_{max}+I_{min}}$) of $ 70 \pm 2 \% $. By analyzing the light scattered from the end of the WG in figure \ref{pol} (e), we find that the light is polarized along the WG axis with a $DoP=87 \pm 3 \%$. Both excitation and detection efficiencies are both highest when the electric field is along the WG axis. For our measurement geometry we calculate, that the combined excitation and detection efficiency is $26 \pm 3\%$ relative to the a co-polarized geometry where excitation and detection polarizations are aligned along the WG axis. This is in good agreement with a simple model using three linear polarizers, whereby two are perpendicular to each other and the third is tilted by a specific angle in between. In this geometry, the intensity measured after the third polarizer is at best $25\%$ of the initial intensity, if the second polaizer is tilted by $45^\circ$. In our experiment, the second polarizer can be asociated with the waveguide, since we excite a mode propagating along the waveguide axis.

%
%
\begin{figure}[t!]
\includegraphics[width=0.95\columnwidth]{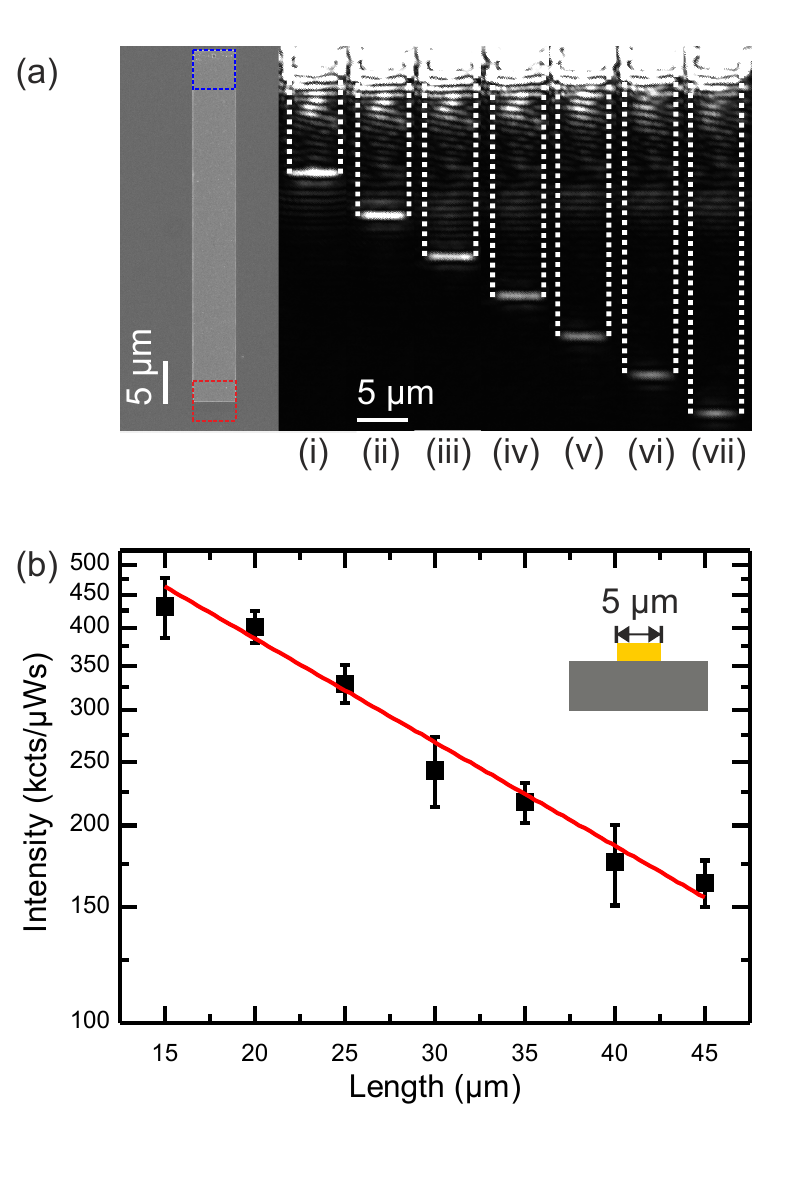}
\caption{\label{meas} (a) left: SEM image of a $5~\mu m$ wide and $45~\mu m$ long WG. Right: a series of images obtained using a CCD camera. (i) a $5~\mu m$ wide and $15~\mu m$ long WG to (vii) $5~\mu m$ wide and $45~\mu m$ long WG. The intensity of the scattered light clearly follows waveguide end. (b) A logarithmic plot of the intensity as a function of the waveguide length. Fitting the data with an exponential decay yields the propagation length of a $5~\mu m$ wide WG excited with $\lambda=830~nm$. Inset: cross section of a Au-waveguide on GaAs }
\end{figure}

%
%
One key figure of merit of the SPP waveguides is the propagation length $L_{SPP}$. To directly measure $L_{SPP}$ for our samples, sets of five nominally identical WGs were fabricated and investigated for $L_{WG}=15-45~\mu m$ and $w_{WG}=2-5~\mu m$. The intensity of the transmitted light is maximized by exciting the in-coupling hole and detecting the integrated intensity of the pixels around the remote end of the WG, denoted by the detection rectangle of figure 1c. Typical integration times and excitation power densities are $t_{int}=1~s$ and $P = 650~{W {cm}^{-2}} $, respectively. The integrated intensity is normalized for the power of the laser and integration time to remove fluctuations of the laser power. A typical set of images recorded for different $L_{WG}$ and a common width of $w_{WG}=5~\mu m$ is presented in figure \ref{meas} (a). The position and intensity of the scattered light from the waveguide end are clearly correlated to the length: As $L_{LG}$ increases from $15 \mu m$ in (i) to $45 \mu m$ in (vii), the position of the out-coupling clearly tracks the position of the end of the WG. With increasing length, the normalized intensity reduces from $430 \pm 45~\frac{kcts} {\mu W s}$ in (i) to $164 \pm 14 ~\frac{kcts} {\mu W s}$ in (vii). We measured 5 WGs with nominally identical $L_{WG}$ and $w_{WG}$ to obtain statistics, independent of microscopic structural differences that influence the absolute in- and out-coupling efficiencies. After calculating the mean and the standard deviation for a particular set of WGs, we plot the intensity on a logarithmic scale as a function of the waveguide length. An example of the results obtained using this analysis procedure is presented in figure \ref{meas}(b). We obtain a good fit to the data in figure \ref{meas}(b) using an exponential decay from which we extract the SPP propagation length $L_{SPP}$. For the $5~\mu m$ wide waveguide presented in figure \ref{meas} we obtain $L_{SPP}=27\pm 2~\mu m$ when the system is excited at $\lambda = 830~nm$. This value is in very good agreement with analysis of leaky SPP modes of metallic WG on glass substrate presented in athe literature\cite{Flynn2010}, suggesting that the SPP mode propagates at the Air-Au interface. This conclusion is supported by wavelength dependent measurements presented below.

\begin{figure}[t!]
\includegraphics[width=0.95\columnwidth]{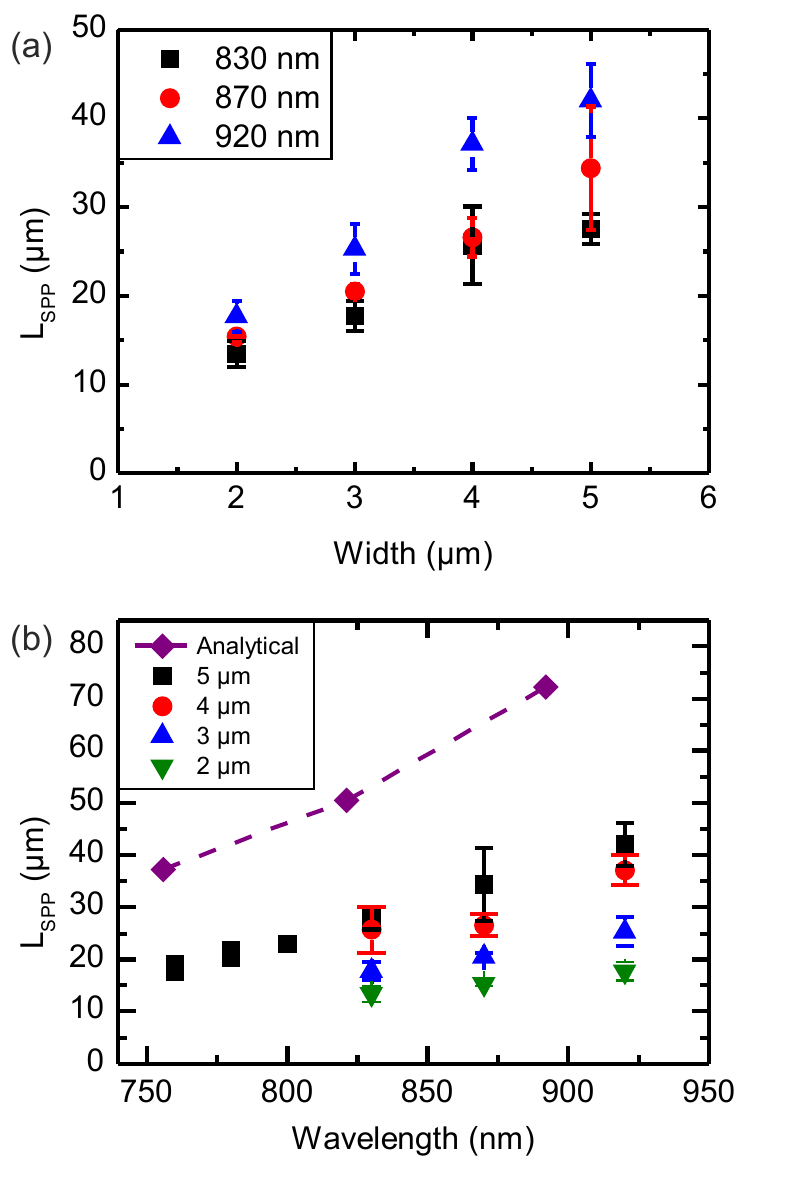}
\caption{\label{prop} (a) Propagation length measurements performed for WG-widths from $2~\mu m$ to $5~\mu m$: For all excitation wavelengths we find that a broader WG-width increases the propagation length. (b) Plot of the propagation length as a function of the excitation wavelength: The propagation length of a $5~\mu m$ wide waveguide has been measured from 760nm to 920nm. The data reproduces the theoretically expected trend to increase with longer wavelengths. }
\end{figure}

%
%

As a step towards sub-wavelength structures we performed similar studies on WGs with widths of $w_{WG}=$ 5, 4, 3 and $2~\mu m$ for $\lambda_{free}=830~nm$, 870~nm and 920~nm on either side of the GaAs band edge. Furthermore, we investigated exemplary the wavelength dependence of $w_{WG}=5~\mu m$ from $\lambda_{free}=760~nm$ to 920~nm. An overview of these measurements is presented in figure \ref{prop} (a); $L_{SPP}$ is plotted as a function of the WG width. For all wavelengths investigated the $L_{SPP}$ reduces for smaller $w_{WG}$. For example, the propagation length reduces from $34 \pm 7~\mu m$ for $w_{WG}=5~\mu m$ to $15\pm 1 ~\mu m$ for $w_{WG}=2~\mu m$ at $\lambda=870 nm $. The pronounced reduction of  $L_{SPP}$ arises from a reduced number of supported SPP modes,as suggested in previous studies employing near-field measurement techniques \cite{{weeber2003},{zia2005}}. This observation indicates, that for sub-wavelength WGs $L_{SPP}$ is likely to be reduced to a few micrometer or less.
Figure 3 (b) depicts $L_{SPP}$ as a function of the wavelength revealing a reduction of $L_{SPP}$ from $42\pm 4 \mu m$ to $18	\pm 2 \mu m$ as the wavelngth is reduced from $\lambda_{free}=920nm$ to 760~nm for a $w_{WG}=5~\mu m$. The violet curve plotted in figure \ref{prop} (b) shows a calculation $L_{SPP}$ at the Au-Air interface obtained using the dispersion relation\cite{novotny2006} for an infinite metal film  and material properties of gold in the near infrared \cite{johnson1972},

\begin{equation}
L_{SPP} \approx \sqrt{\frac{\epsilon_1' +\epsilon_2}{\epsilon_1' \epsilon_2 }}  \frac{\epsilon_1' (\epsilon_1' +\epsilon_2)}{\epsilon_1'' \epsilon_2 } \frac {\lambda}{2 \pi}
\label{prop-ana}
\end{equation}

Although the measured values lie typically $40-60 \%$ below these values, the wavelength dependence is clearly qualitatively reproduced by eqn. \ref{prop-ana}. The same behavior can be observed in narrow waveguides, as shown by figure \ref{prop} (b). The smaller values of  $L_{SPP}$ measured in our experiment arises from the finite width of the waveguide and the polycrystalline nature of the Au-film that introduces losses due to scattering\cite{ditlbacher2005silver}. The polycrystallinity was investigated atomic force and scanning electron microscopy, revealing  grain sizes of $20 \pm 10 nm$. Furthermore, the observed behavior proves that the plasmon propagation at the Air-Au-interface, since no strong reduction  of $L_{SPP}$ is observed for wavelengths corresponding to photon energies above the GaAs bandgap.

%
%
In summary, we have reported a systematic characterization of the propagation of surface plasmon polaritons along few micrometer wide Au waveguides defined lithographically on GaAs. Polarization dependent measurements revealed that SPPs can be generated by exciting and detecting along the WG axis with $DoP_{exc}= 70 \pm 2 \% $ and $DoP_{det}=87 \pm 3 \%$, in excitation and detection, respectivily. Using a 45$^\circ$ tilted geometry we excite and image SPPs while suppressing the exciting laser by $10^4$. The propagation length of different waveguides was determined using different wavelengths for the excitation: Increasing $w_{WG}$from $2~\mu m$ to $5~\mu m$  increases $L_{SPP}$  from $15\pm 1 ~\mu m$ to $34 \pm 7~\mu m$ using an excitation wavelength of $\lambda=870 nm $. Furthermore, we find, that the excitation wavelength has a strong influence on $L_{SPP}$: Increasing the wavelength from $760~nm$ to $920~nm$ increases the propagation length from $18	\pm 2	\mu m$ to $42	\pm 4 \mu m$ on a $5~\mu m$ broad WG. The deterministic control of the position and shape of the plasmonic structures by means of electron beam lithography combined with near surface self-assembled InGaAs/GaAs quantum dots promises efficient on-chip generation and guiding of single plasmons for future applications in nano-scale quantum optics.

\begin{acknowledgments}
We acknowledge financial support of the DFG via the SFB 631, Teilprojekt B3, the German Excellence Initiative via NIM, and FP-7 of the European Union via SOLID. The author gratefully acknowledges the support of the TUM Graduate School's Faculty Graduate Center Physik at the Technische Universit\"at M\"unchen.
\end{acknowledgments}

\bibliography{ref}

\newpage
\begin{appendix}
\section{Methods}

The samples investigated were defined on undoped GaAs wafers. After cleaving the wafer the sample is cleaned using acetone in an ultrasonic bath and is rinsed with isopropanol (IPA). A bi-layer resist (Polymethylmethacrylat 200K AR-P 641.07) is coated at 4000~rpm  for 40~s and is baked at 180~$^\circ$C for 10 min, producing a resist thickness of $365 \pm 5 nm$. After cooling, PMMA 950K (AR-P 671.02) is spin-coated at 4000~rpm for 40~s, baked for 6~min at 180~$^\circ$C, to produce a bi-layer resist with a total thickness of $460\pm10~nm$. A  JEOL JSM-6400 with a Raith interferometric table is used for the electron-beam lithography. The dose used for the exposure was $490 \pm 25~\mu As/cm^2$. After the electron beam writing the sample is developed in Methylisobutylketon (MIBK) diluted with IPA (1:3) for 30~s. To stop the development the sample is rinsed with pure IPA. For the metalization a electron beam  evaporator is used to deposit 100~nm of gold at a low rate of 1~\AA/s to produce a uniform film with an RMS roughness of $2\pm1~nm$. 

\end{appendix}

\end{document}